# An Experimental Study of the Acoustic Field of a Single-Cell Piezoelectric Micromachined Ultrasound Transducer (PMUT)

Bibhas Nayak
Centre for Nano Science & Engineering
Indian Institute of Science
Bangalore, India
bibhasnayak@iisc.ac.in

Harshvardhan Gupta
Centre for Nano Science & Engineering
Indian Institute of Science
Bangalore, India
gharshvardha@iisc.ac.in

Kaustav Roy
Centre for Nano Science & Engineering
Indian Institute of Science
Bangalore, India
kaustav@iisc.ac.in

Anuj Ashok
Centre for Nano Science & Engineering
Indian Institute of Science
Bangalore, India
anujashok@iisc.ac.in

Vijayendra Shastri
Centre for Nano Science & Engineering
Indian Institute of Science
Bangalore, India
antonya@iisc.ac.in

Rudra Pratap
Centre for Nano Science & Engineering
Indian Institute of Science
Bangalore, India
pratap@iisc.ac.in

*Abstract*—Piezoelectric micromachined ultrasound transducers (PMUTs) have gained popularity in the past decade as acoustic transmitters and receivers. As these devices usually operate at resonance, they can deliver large output sound pressures with very low power consumption. This paper explores the influence of the transmitter's packaging on the radiated acoustic field in air. We run simplified axisymmetric numerical models to observe the change in the acoustic field and directivity with respect to the device's package dimensions. The simulations demonstrate a notable change in the directivity of transmitter based on the size of the baffle. Experimental measurements are carried out to validate the simulations, which can prove useful in designing packages for transmitters to meet application specific requirements.

## I. Introduction

A piezoelectric micromachined ultrasound transducer (PMUT) is typically a multilayered MEMS structure, fabricated by deposition and patterning of its constituent piezo- electric and electrode films on a structural layer, followed by the release of its vibrating element, generally a diaphragm by backside etching [1]. Its piezoelectric nature allows it to be used for both transmitting and receiving ultrasound. The operating frequencies of PMUTs generally range from a few tens of kHz to hundreds of MHz. With a small footprint, low operating voltage and low power draw, PMUTs find use in applications such as medical imaging [2], photoacoustics [3]–[6], range finding [7], nondestructive evaluation [8], fingerprint sensing [9], density sensing [10]–[14] and data-over-sound [15].

The performance requirements from the transducer are generally dictated by the application. Apart from frequency response, output pressure and sensitivity, the directivity of the transducer can also influence its performance. For applications such as range-finding, a highly directive, on-axis radiation pattern is desirable. In contrast, proximity sensing and data-over-sound applications may benefit from a wider or omnidirectional pattern.

The packaging of a single PMUT is not a widely explored area in the field of ultrasound. There has been some work on widening the directivity of piston-like ultrasonic transducers with 3D printed baffles [16]. Other studies deal with generating highly directional ultrasonic radiation from stepped circular plates [17]. Additionally, the field of audio reproduction has generated a fair amount of literature on the shape and size of speaker cabinets [18]–[20] in which the dimensions of the speaker cabinet arguably play an important role in the directivity and frequency response at a desired listening position.

The cause behind the aforementioned influence on directivity and frequency response is attributed to the interference caused by diffraction of sound around the speaker cabinet [21]. In particular, it is the diffraction of sound from the frontal edges of the cabinet that can influence the radiated acoustic field significantly. The interference introduces changes to the frequency response at the listener's position and this is often regarded as unwanted coloring of the sound. It simultaneously modifies the radiation pattern of the speaker in the resulting sound field. This is one of the reasons why some high-fidelity audio systems have complex or unconventional shapes for their speaker cabinets.

In this paper, we study the effects of the package dimensions on the radiation pattern of a PMUT coupled with air. Using finite element simulation of a coupled electrical-mechanical-acoustical model of a PMUT in air, we visualize the acoustic field generated for three different baffles. The acoustic field results are validated using experimental measurements using a reference microphone and a field scanning setup.

## II. THEORY

The following sections discuss the possible factors that influence the radiation pattern of a speaker. We first look at the directivity of a vibrating circular piston on an infinite baffle and then discuss the phenomenon of diffraction. The presence of diffracted waves and how they can cause interference are explained with the help of simple geometrical concepts.

### A. Radiation from a circular piston on an infinite baffle

The model of a piston mounted on a rigid infinite baffle is often used for its simplicity in understanding the directivity of a loudspeaker, especially due to the piston-like motion of the conventional loudspeaker diaphragm [19]. The directivity $D(\theta)$ of a circular piston represents its radiation across different angles of observation. It is mathematically written as

$$D(\theta) = \frac{2J_1(ka\sin(\theta))}{ka\sin(\theta)}, \quad (1)$$

where $J_1$ is the Bessel function of the first kind, $k$ is the angular wavenumber of radiated waves, $a$ is the radius of the piston and $\theta$ is the angle of observation [22].

However, this expression faces problems with raised baffles on an infinite plane, discussed in Sec. III, where diffraction effects alter directivity due to interference from newly formed secondary sources along the baffle's edges.

### B. Diffraction of Sound

Diffraction is understood as the bending of travelling waves upon interaction with a discontinuity or obstacle that has dimensions comparable to their wavelength [23]. According to Huygens principle, diffraction causes the obstacle to act as a secondary source of spherical waves. In everyday scenarios, this phenomenon enables a person talking behind a tree or the bends of a hallway to be audible over the said obstacle. In our case, the edges of the die and printed circuit board (PCB) of the PMUT can act as new secondary sources. Depending on their dimensions, these secondary sources can interfere with the primary acoustic radiation from the transducer, influencing its directivity and sound field.

### C. Baffle shape

The baffle's shape determines the resulting path differences between direct and secondary sources. Due to diffraction, we may consider the edges of the baffle to be composed of a large number of infinitesimally small secondary sources [24]. For this discussion, we will use the geometry of a simple circular baffle to help explain the effect of baffle diffraction [25]. If we assume a radiating monopole placed at the centre of the baffle, there is a constant path difference $x_p$ between two groups of waves reaching a point $R$ on the central axis (see Fig. 1):

1) Waves that arrive directly from $S$.
2) Waves that arrive from $S$ after having diffracted from the baffle's edge $B$.

A sketch illustrating the paths of the direct and diffracted waves to an on-axis receiver using a circular baffle, is shown in Fig 1). Consider a radiating monopole situated at $S$, with the secondary sources $B$ being formed at the circumference. Since the path difference $x_p$ (see Fig. 1) is constant, a circular baffle will offer the strongest interference effects with $x_p$ being a function of the radiated wavelength. This occurs due to a delayed copy of the same signal arriving at $R$ from $B$. Other baffle shapes will reduce the severity of such effects as the path difference is no longer held constant with respect to all points along their edges. It follows that adding a 3D geometry (such as a spherical or trapezoidal baffle) will greatly lessen the interference effects at $R$ and in the radiated sound field [18].

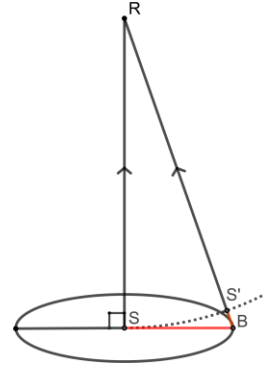

Fig. 1: Direct and diffracted waves arriving at $R$ from the primary source at $S$ and secondary source at $B$, respectively with a path difference $x_p$, where $x_p = SB + BS'$ (highlighted in red).

### D. Path difference

In Fig. 1, the on-axis receiver $R$ is at a distance $RS$ (height) and $RB$ (hypotenuse), from the monopole and the baffle edge, respectively. For this example, $R$ remains equidistant from $B$ due to the circular geometry.

A method of determining $x_p$ between the sound waves reaching $R$ is to employ reciprocity theorem [26]. The theorem states that the pressure observed at a given receiver due to some acoustic source, would remain the same if their respective positions were swapped. From Fig. 1, the wavefront from $R$ (denoted by dotted arc $SS'$) approaching the baffle's edge would need to travel an additional $BS' + SB$ before it reaches $S$.

Assuming spherical waves being generated, it follows that the path difference is,

$$x_p = SB + BS' = SB + RB - RS, \quad (2)$$

where,

$$RB = (SB^2 + RS^2)^{0.5} \quad (3)$$

For in-phase addition of waves (constructive interference) to occur at $R$, the path difference has to be an odd multiple of half times the radiated wavelength [27],

$$x_p = (2n+1)\frac{\lambda}{2}, \quad \text{where } n = 0, 1, 2, \ldots \quad (4)$$

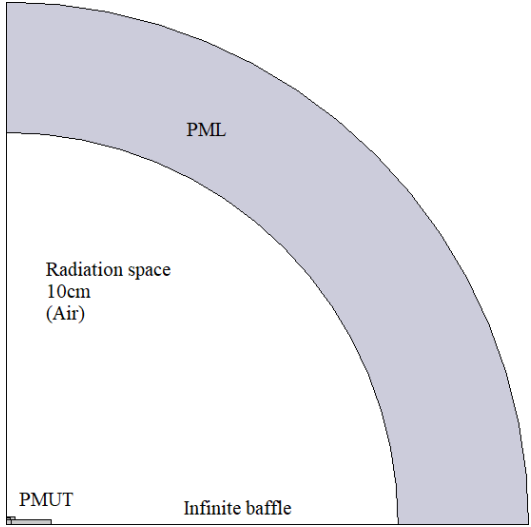

Fig. 2: 2D axisymmetric FEM model for acoustic radiation from a PMUT with the die and PCB placed on an infinite baffle.

TABLE I: Material parameters used for simulating the PMUT.

| Parameter | Si | Pt | PZT | Au |
|---|---|---|---|---|
| Density (kg/m$^3$) | 2329 | 21450 | 7500 | 19300 |
| Young's modulus (GPa) | 170 | 168 | 110 | 70 |
| Poisson's Ratio | 0.28 | 0.38 | 0.39 | 0.44 |

For out-of-phase cancellation of waves (destructive interference) to occur at $R$, the path difference has to be an integral multiple of the full radiated wavelength [20],

$$x_p = (n + 1)\lambda, \qquad \text{where } n = 0, 1, 2, ... \quad (5)$$

## III. SIMULATION

To observe the influence of baffle diffraction on the directivity of the PMUT, a 2D axisymmetric finite element model (FEM) of the PMUT was simulated in COMSOL. The following subsections detail the simulation model and the results.

### A. Simulation setup

Figures 2 and 3 illustrate the geometry of the FEM model. In Fig. 2, the PMUT along with its package is placed on an infinite baffle, radiating into a space of $2\pi$ steradian with axisymmetry. The radiation space of 10 cm terminates into an absorbing, perfectly matched layer (PML) which simulates an anechoic environment. Figure 3 shows the PMUT placed on a PCB, in which the device layer is 10 µm thick with a radius of 1 mm. On the device layer lies a 650 nm thick PZT layer that has a residual tensile stress of 300 MPa. The handle layer which serves as the die's baffle is 650 µm thick. The PMUT is placed over a PCB that is 1 mm thick and contains a 1.5 mm radius hole underneath the PMUT. The material parameters used for the PMUT are defined in the Table I.

In our simulation, the die's baffle diameter is kept fixed at 4.5 mm while the PCB's baffle diameter is varied to observe changes in the radiation pattern.

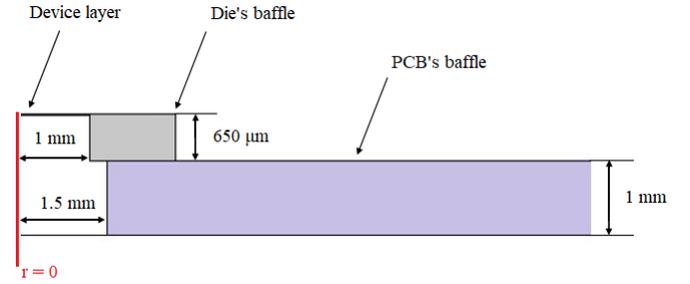

Fig. 3: Close up view of 2D axisymmetric PMUT on its PCB, with the reference axis indicated by $r = 0$.

### B. Results

An eigenfrequency study was carried out on the PMUT's diaphragm, which returned a fundamental resonant frequency of 50.7 kHz. This was followed by a frequency domain study at resonance for a reference case where the PMUT's diaphragm was flush with the infinite baffle. For observing the influence of the package, the PMUT was simulated with two different PCB baffles of diameters 8 mm and 24 mm, respectively. These diameters are equal to the breadths of the PCB baffles used in our measurements, wherein the PMUT was placed on a 12 mm x 8 mm and a 24 mm x 24 mm PCB baffle respectively.

Figure 4 depicts the sound field when the PMUT is used with an infinite baffle. Its radiation appears to be very omni-directional with no visible pressure lobing and this is due to the lack of any diffracted waves which would have caused interference. Figure 5 represents the acoustic field of the PMUT with a PCB baffle diameter of 8 mm. A strong singlelobe is observed on-axis (the vertical axis here). With a larger baffle diameter of 24 mm as shown in Fig. 6, a diminished main lobe is generated with three sidelobes around it.

The directivity of the PMUT under these different conditions is presented in Fig. 7. The behaviour of the PMUT with its diaphragm flush with an infinite baffle is quite similar to the directivity as predicted by Eq. (1). In other cases, the influence of diffraction on PMUT's directivity is quite noticeable.

## IV. EXPERIMENTAL MEASUREMENT

### A. Measurement setup

A 3D printer was re-purposed to behave as a computer-controlled 3-axis motion system for positioning a GRAS 46DP-1 reference microphone. The microphone was powered using a B&K Type-2829 microphone power supply. The device under test was placed on the bed of the 3D printer and driven at $1\text{V}_{\text{rms}}$ using a Zurich Instruments MFLI lock-in amplifier that also measured the frequency response of the transducer. This allowed for physically mapping the sound field of the given source by sampling the pressure in a grid-like pattern in front of it. It was ensured that the spatial sampling resolution, or the step size of the printing head's movement was less than $\lambda/6$, where $\lambda$ is the wavelength of the radiated waves. This was done so as to reproduce the scanned acoustic field with minimal artifacts in the final image.

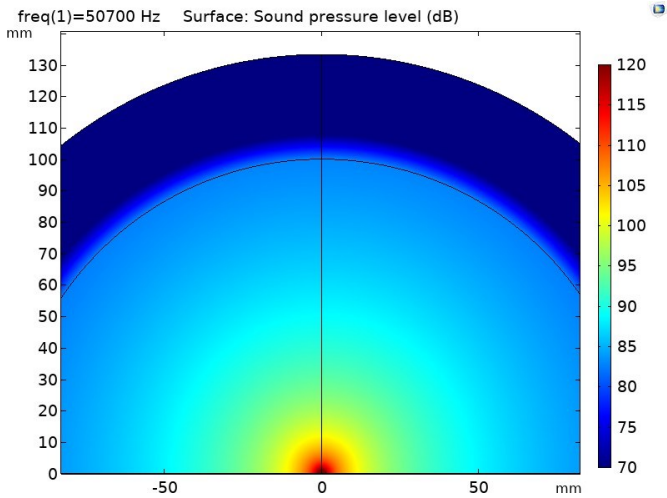

Fig. 4: Simulated acoustic radiation at 50.7 kHz with infinite baffle and with PMUT diameter of 2 mm.

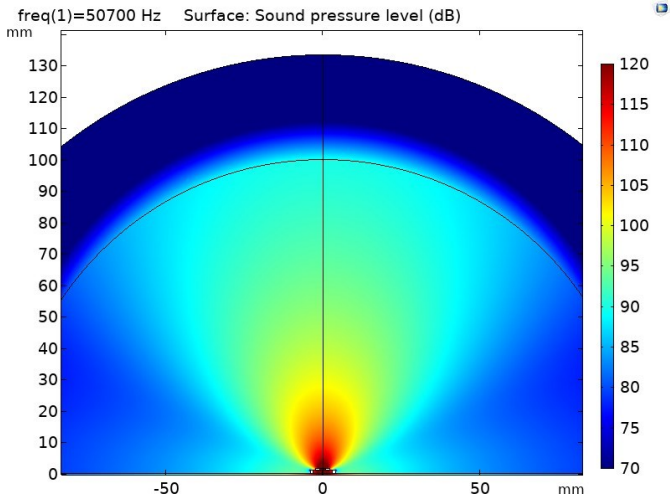

Fig. 5: Simulated acoustic radiation at 50.7 kHz with baffle diameter of 8 mm and PMUT diameter of 2 mm.

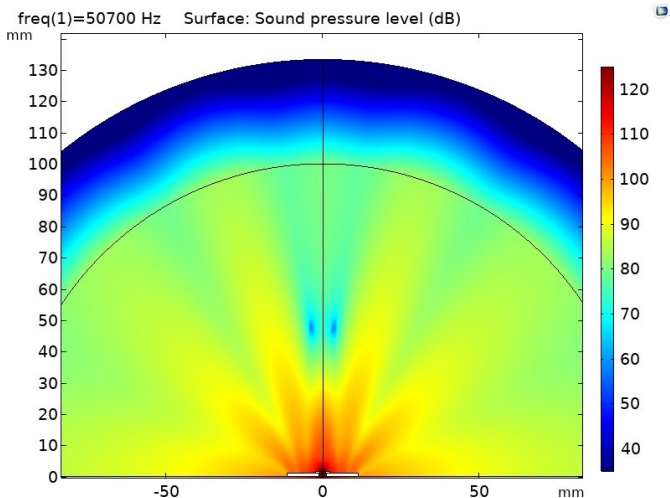

Fig. 6: Simulated PMUT's radiation at 50.7 kHz with baffle diameter of 24 mm and PMUT diameter of 2 mm.

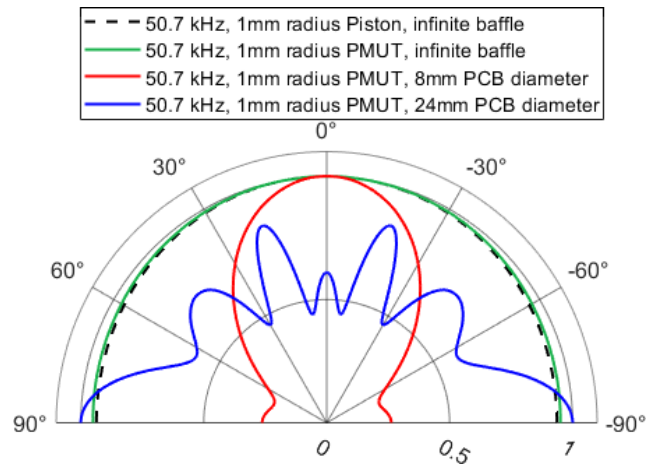

Fig. 7: Directivity of the PMUT with different baffle radii. The piston-on-infinite-baffle is presented as an analytical comparison.

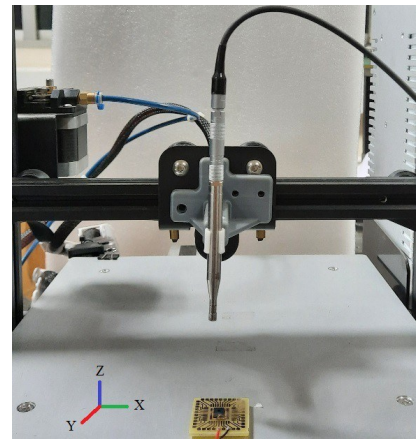

Fig. 8: 3D printer converted into a 3-axis motion system with a microphone arm, used for mapping the acoustic field of the PMUT (in picture, on a 24 mm x 24 mm PCB baffle).

The measurement setup is shown in Fig. 8 where the PMUT is placed on the 3D-printer's bed, or the CartesianXY plane. The acoustic field scan was performed in the XZ plane. The PMUT was oriented such that the breadth of its PCB faced the Y direction. The motion system was controlled using a LabVIEW program running on a Windows PC which also recorded the data from the lock-in amplifier. Using the measured data, acoustic field maps were plotted where the colour-bar indicates the sound pressure level in decibels.

*B. Results*

Using the measured scan data, the radiated acoustic field of the PMUT was plotted (Figs. 9 and 10). Although minor differences can be observed between the measurements and the simulation results due to the simulated model being axisymmetric, the acoustic field maps obtained from the measurement are in good agreement with the behaviour observed in the simulation (Figs. 5 and 6), where the smaller PCB showsa prominent central lobe in comparison to the larger PCB

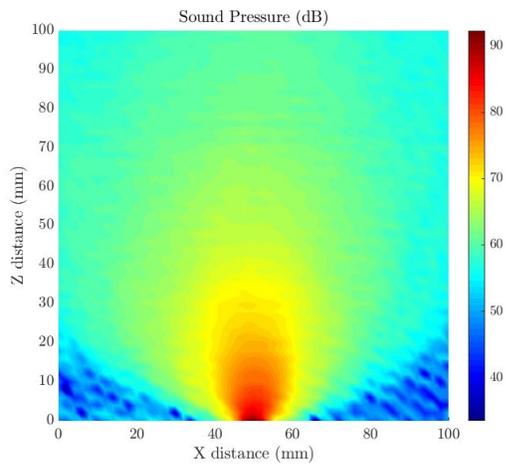

Fig. 9: Measured acoustic radiation at 50.7 kHz with PCB dimensions of 12 mm x 8 mm and PMUT diameter of 2 mm.

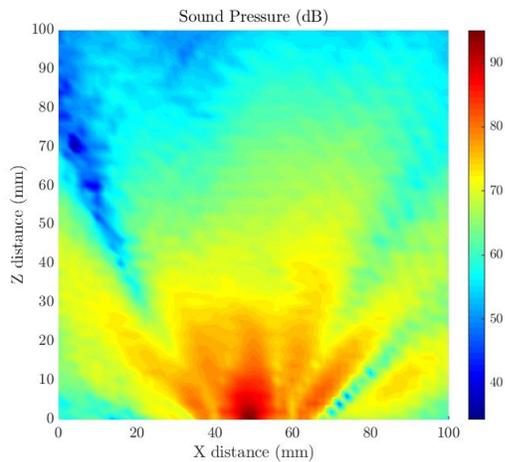

Fig. 10: Measured acoustic radiation at 50.7 kHz with PCB dimensions of 24 mm x 24 mm and PMUT diameter of 2 mm.

which indicates the presence of multiple sidelobes caused by diffraction at the edges of the PCB.

The discrepancy in the magnitude of the sound pressure levels between the simulation and measurement is due to the simulation running with COMSOL's default electro-mechanical coupling parameters for PZT. A comprehensive evaluation of the material parameters of the PZT film used in the fabricated PMUTs was not conducted prior to performing the simulation for this paper.

## V. CONCLUSION

Single PMUTs are almost never packaged on an infinite baffle, and are always diced to a finite die size, and bonded to a PCB package. Diffraction effects from the edges of the die and the package can significantly influence the directivity pattern of the PMUT, which has been demonstrated using acoustic field simulations and validated using measured scans of the sound field. This preliminary work highlights the importance of the package for a PMUT, which can be engineered towards application-specific requirements to improve the directionality or to make the radiation pattern more omnidirectional.